\documentclass[10pt]{report}
\usepackage{graphicx}
\usepackage{amsfonts}
\usepackage{float}
\usepackage{caption}
\usepackage{pdfpages}
\usepackage{multicol}
\usepackage{amsmath}
\usepackage{etoolbox}
\usepackage[margin=0.8in]{geometry}
\usepackage{cite}
\patchcmd{\thebibliography}{\chapter*}{\section*}{}{}

\begin{document}

\begin{center}\textbf{\Large{Revisiting the distance duality relation using a non-parametric regression method }}\end{center}
\begin{center} {\large Akshay Rana$^{a1}$, Deepak Jain$^b$, Shobhit Mahajan$^a $ and Amitabha Mukherjee}$^a$\\
$^a$Department of Physics \& Astrophysics, University of Delhi, Delhi 110007, India\\
$^b$Deen Dayal Upadhyaya College, University of Delhi, New Delhi 110015, India

$^1$Email- arana@physics.du.ac.in
\end{center}

\textbf{\large{Abstract:}} The interdependence of luminosity distance, $D_L$  and angular diameter distance, $D_A$ given by the distance duality relation  (DDR) is very significant in  observational cosmology. It is very closely tied with the temperature- redshift relation of Cosmic Microwave Background (CMB) radiation. Any deviation from $\eta(z)\equiv \frac{D_L}{D_A (1+z)^2} =1$ indicates a possible emergence of  new physics. Our aim in this work is to check the consistency of these relations using a non-parametric regression method namely,  LOESS with SIMEX. This technique avoids dependency on the  cosmological model  and works with a  minimal set of assumptions. Further, to analyze the efficiency of the methodology, we simulate a dataset of $200$ points of $\eta (z)$ data based on a phenomenological model $\eta(z)= (1+z)^\epsilon$. The error on the  simulated data points is obtained   by using  the temperature of CMB radiation at various redshifts. For testing the distance duality relation,  we use the JLA SNe Ia data for luminosity distances, while the angular diameter distances are obtained from radio galaxies datasets. Since the DDR is linked with CMB temperature - redshift relation, therefore we also use the CMB temperature data to reconstruct $\eta (z)$. It is important to note that with CMB data, we are able to study the evolution of DDR upto a very high redshift  $ z = 2.418$. In this analysis, we find no  evidence of deviation from $\eta=1$  within a $1\sigma$ region in the entire redshift range used in this analysis ($0 < z \leq 2.418$).

Keywords: Distance duality relation, CMB Temperature, Radio Galaxies , LOESS, SIMEX.

\vspace{0.3cm}

\begin{flushleft}\textbf{\large{1. Introduction}}\end{flushleft}
 The relation between the luminosity distance $D_L$ and the angular diameter distance $D_A$ was derived by Etherington in 1933 \cite{ethr}. This relation is known as the distance duality relation (DDR).
 \begin{equation} D_L=  D_A (1+z)^2\end{equation} It is valid for all curved space-times and  holds as long as  (i) Photon number is conserved, (ii) Gravity is described by a metric theory and (iii) Photons travel along null geodesics \cite{ellis}. We can define a parameter $\eta(z)$ as
\begin{equation} \label{etadef}
\eta(z) \equiv \frac{D_L}{ (1+z)^2 D_A}=1
\end{equation}
Then, any deviation from the DDR will give us $\eta(z) \neq 1$. \\

The DDR  is of fundamental significance in observational cosmology. It plays an important role in the  observations linking galaxy clusters  \cite{lima1,lima2} and gravitational lensing \cite{schneider}. It also allows us to derive the  proportionality between the temperature of CMB photons $(T_{CMB})$ and their redshifts \cite{komatsu}.
 \begin{equation} \label{temprad}
  T_{CMB}(z)=  T_0 (1+z)\end{equation}
where $T_0$ is the average CMB temperature today. This relationship is not restricted to any particular metric theory and it holds in the general theory of relativity as long as photon number is conserved and photons follow null geodesics. Hence the violation of the CMB temperature - redshift relation is also related to the violation of DDR. Deviation from both of these relations points to a violation of one or more of the above mentioned conditions and therefore highlights the importance of checking the consistency of different datasets as well as the cosmological models. Many observational datasets have been used to check the consistency. For instance,  radio galaxies \& ultra compact radio sources\cite{basset}, X- ray gas mass fraction of galaxy clusters\cite{rfl,goncalves, goncalves22 }, Cosmic Microwave Background\cite{ruth,  gfr, sysky}, Gamma Ray Bursts\cite{rfl2},  H 21 cm signal from disk galaxies\cite{khedekar} and Baryon Acoustic Oscillation (BAO) \cite{cong, cardone} are among the observations used to check this important relation.

Many authors have also used DDR to constraint the cosmic opacity of the universe \cite{more, avgo,nicole, djain}.

Several theoretical efforts have been made to study violations of DDR and the possible  emergence of new physics. For instance, supernova dimming due to the cosmic dust  affects the luminosity distance measurement \cite{csaki,jas,suto,pier} and  photon-axion conversion in intergalactic magnetic fields violates photon conservation\cite{mirizzi,avgo} and hence could cause a  deviation from  $\eta = 1$.  Similarly, though DDR should hold  under the assumption of a  metric theory, Piazza \& Schucker (2016) try to understand the behaviour of DDR in a non-metric theory of gravity \cite{mincos}.

 The violation of DDR can be studied broadly in two ways. First, we can analyze   DDR in the framework of different cosmological models which  can explain the present accelerated
  expansion of the universe \cite{marek}.  Alternatively, we could study it in a model independent way. Since the present cosmological model of the universe is not known, a model independent methodology is probably a  better approaches to  analyze  DDR.  The  model independent approach can further be subdivided into two different branches: parametric and non-parametric.

  Several different forms of parametrization of $\eta(z)$ have been used in the literature\cite{remya ,holanda, holandaI ,meng ,liang ,wu  ,yang}. This phenomenological approach has the disadvantage that the final result may depend upon the assumed form of  $\eta(z)$.  Therefore  non-parametric methods seem to be a robust way to extract information from the data. Gaussian Process (GP), Genetic Algorithm (GA)  and  Smoothing Method (SM) techniques have been used  to study the  variation of $\eta(z)$ or to check the consistency of  different datasets in a non-parametric way\cite{costa,remya1,zheng,sahni,nesseries}. However, both GP and SM  methods require an initial prior or cosmological guess model, which may lead to a biased result. On the other hand, errors may be underestimated in the GA method \cite{nesseries}.

In this work, we use a  robust, strong and computationally easy non-parametric regression technique called \textbf{LO}cally w\textbf{E}ighted \textbf{S}catterplot \textbf{S}moothing (LOESS) to test the consistency of  DDR\cite{cleveland1,cleveland2}. Without assuming any prior or cosmological model,  LOESS recovers the global trend of data by studying the appropriate number of neighbouring data points around the focus point.
 Further, to take measurement errors of data into account, we use \textbf{SIM}ulation and \textbf{EX}trapolation method (SIMEX)\cite{cook1,cook2,biewen}. In Cosmology, this technique has been used in detail  by Montiel et al for reconstruction of the  cosmic expansion history \cite{montial}. Recently this methodology has also been used to reconstruct the Om diagnostic \cite{fabris} and to constrain the transition redshift \cite{nisha}. The paper is organized as follows. In Section 2, we discuss the datasets (both real and simulated ) used in this work. In Section 3, we briefly summarize the LOESS and SIMEX techniques. We describe results in Section 4. Finally we end with a discussion in Section 5.

\begin{flushleft}\textbf{\large{2. Datasets}}\end{flushleft}

 \textbf{2.1 Dataset A }: In order to obtain the observed value of $\eta$, i.e. $\eta_{obs}$,  we need datasets of luminosity distance and angular diameter distance. $D_L$ is taken from JLA dataset of  $740$ supernovae in the redshift interval $0.01<z<1.3$  \cite{betoule}. SNe Ia Data is given in the form of distance modulus $\mu$  along with the uncertainty $\sigma_{\mu}$. By using the relation $\mu= 5log_{10}D_L +25 $, we get the luminosity distance $D_L$ and the corresponding variance $\sigma_{D_L}^2$.
To determine the angular diameter distance $D_A$, we use the dimensionless coordinate distance of 20 FRIIb radio galaxies  in the redshift range $0.056<z< 1.8$  \cite{routhdaly} .

However,  to calculate  $\eta_{obs}$, we need both $D_L$ and $D_A$ at the \textbf{same} redshift. In  an earlier work, a moderate redshift criterion, $\Delta z<0.005$ (where $\Delta z= |z_{galaxy}-z_{supernova}|$) was adopted and the nearest SNe Ia was selected for each radio galaxy \cite{remya ,wu}. But selecting only one SNe within the same redshift range  usually gives rise to large statistical errors.  The choice of $\Delta z$ has a crucial impact on the constraints on DDR. Cao \& Liang (2011) used  different redshift binning sizes which varied from  $\Delta z = 0.002$ to 0.005 \cite{cao}. They highlighted how both the choice of $\Delta z$ and number of SNe in each redshift bin may play a very important role in the study of DDR .

In practice $\Delta z$ should not be smaller than 0.002. But this condition is more stringent and may increase the statistical error. Therefore, in order to increase the reliability and accuracy of the test, several  groups have advocated the use of $\Delta z = 0.005$ \cite{holanda,wu}.

In literature, $\Delta z = 0.005$ has an acceptable optimal choice. In this work we use the inverse variance weighted average method \cite{meng}. This is the most suitable choice because it gives the appropriate weight to each  SNe Ia  lying in the region thereby significantly decreasing the statistical error.
 
 The weighted mean value of the luminosity distances for all the Supernovae  in the given region can be written as:

\begin{equation}\label{barDL}
\bar{D}_L = \frac{\sum\limits_{i} \frac{(D_L)_i}{\sigma_i^2}}{\sum\limits_{i}\frac{1}{\sigma_i^2}}
\end{equation}
\begin{equation}\label{barsigmaL}
\bar{\sigma}_L^2= \frac{1}{\sum\limits_ {i} \frac{1}{\sigma_i^2}}
 \end{equation}

where $\bar{D}_L$ is the weighted mean luminosity distance at the corresponding redshift of the  radio galaxy and $\bar{\sigma}_{L}$ is its
uncertainty.

 Finally, we are left with $12$ points of $ \eta_{obs}$ in Dataset A in the redshift  range of $0.056<z<0.996$. The uncertainties in  $\eta_{obs}$ are  calculated using  error propagation.

\vspace{0.5cm}

\textbf{2.2 DDR and CMB temperature - redshift relation} {\bf(Dataset B)}: The distance duality relation and temperature- redshift relation are closely related and have  similar physical consequences. This can be understood by using CMB to measure luminosity distance $D_L$. In general, the  luminosity distance is inversely proportional  to the square root of the observed flux $F$, and directly proportional to the square root of the luminosity $L$ of  the source . It is  interesting that unlike a supernova, galaxy or any other localized source, CMB is a continuous source and has a  blackbody spectrum. Hence, for such sources, we have $L \propto A T_{CMB}^4$ and $F \propto T_0^4$, where $A$ is the area of emission. Now imagine that we are viewing a patch of CMB sky having an angular size $d \Omega$, then the viewing area can be defined in two ways. Firstly, by assuming the CMB sky to be isotropic i.e. by neglecting small anisotropies, then $dA= \frac{d\Omega}{4\pi}A$. Secondly, if we have $D_A$ to be the angular diameter distance then,  $dA=  D_A^2 d\Omega$. On simplifying these relations we obtain \cite{sysky}

\begin{equation}\label{dLdAT}
D_L = \bigg(\frac{T_{CMB}}{T_0}\bigg)^2 D_A\end{equation}

\noindent By using the above equation, one can easily  rewrite $\eta(z)$ in term of temperature of CMB.

\begin{equation} \label{dLdATeta}
\bigg(\frac{T_{CMB}}  {T_0 (1+z)} \bigg)^2\equiv \eta (z)=1 \end{equation}

The observed value of the parameter $\eta_{obs}$ (Dataset B) can be obtained by using the CMB temperature at various redshifts (see Table 1 in \cite{baranov}). Dataset B contains $36$ points in the redshift interval  $ 0 \leq z \leq 2.418$. $13 $ of the data points  are obtained from multi frequency measurements of the  Sunyaev - Zeldovich (SZ) effect and $18$ measurements of $T(z)$ are compiled by using the thermal SZ effect data from the Planck satellite. The rest of  the data points are inferred from the damped Lyman alpha systems and the  fine structure of neutral carbon atoms.

\vspace{0.5cm}

\textbf{2.3 Generation of Mock datasets}: In order to check whether this regression method is accurate enough for reconstructing the cosmological parameters, we use a simulated dataset based on a phenomenological model with realistic errors. The process of simulation can be described  as follows:
\cite{ gon16}. 

1. We assume a  parametrization for $\eta(z)$,  $\eta_{fid}(z)= (1+z)^\epsilon$ , where $\epsilon$ is a constant. For this fiducial model , the  best fit value of $\epsilon$ turns out to be $-0.0319$ on doing  $\chi^2$ analysis with Dataset B.

\begin{figure*}[ht]
\includegraphics[height=6cm,width=8cm,scale=8]{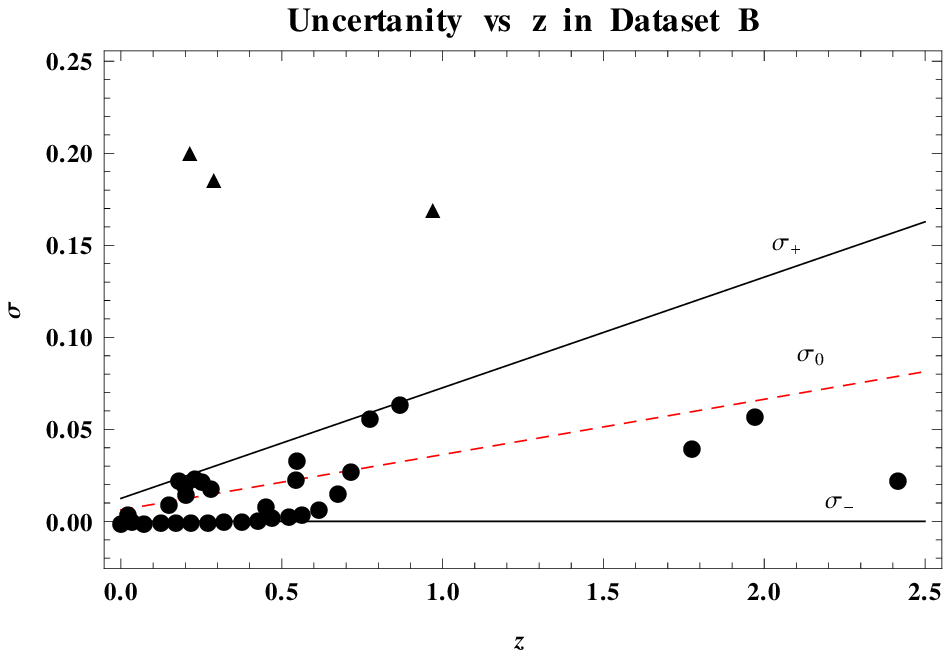}
\hspace{2cm}
\includegraphics[height=6cm,width=8cm,scale=8]{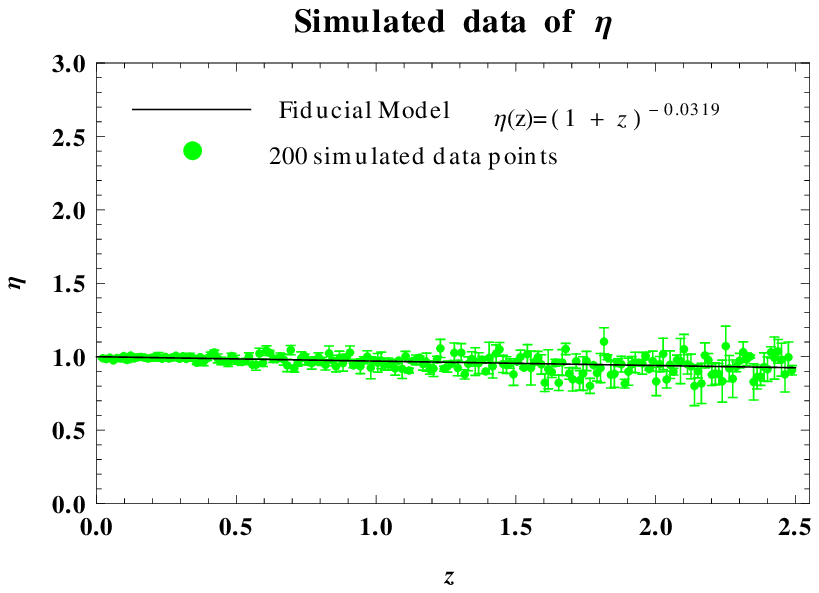}
\caption{\textbf{a)} Left:   Uncertainty of $\eta$ in the CMB temp data (Dataset B). Solid triangles and circles  represent the outlier and non-outlier points respectively. $\sigma_+$ and $\sigma_-$ are  the  upper and lower bounds on error while $\sigma_0$ is the mean estimate of error.  \textbf{b)}   Right: The $200$ Simulated datapoints of $\eta$ generated using realistic errors are shown. Here the black line shows the behaviour of fiducial model.}
\label{fig1}
\end{figure*}

2. The phenomenological approach is followed to find the redshift dependence on the error bars of the simulated data points. To achieve this, we use the real Dataset B to estimate the uncertainties. As shown in Fig \ref{fig1}(a), most of the error bars on the CMB data points ( Dataset B) are contained between the two straight lines, $\sigma_-= 0$ (the horizontal axis) and $\sigma_+= 0.03 z+0.006$ (ignoring a few outliers) . The midline $\sigma_0= 0.015z+0.003$, represents the average uncertainty associated with all future observations. The uncertainty associated with each simulated data point is obtained from the Gaussian distribution centered at the $\sigma_0$ with standard deviation $\frac{\sigma_{+}(z) -\sigma_{-}(z)}{4}$.

3. Using this methodology, we  generate a mock dataset based on realistic errors of the  CMB temperature data. The simulated data points along with their uncertainties are shown in Fig \ref{fig1}(b). The complete details of this procedure are given by Ma \& Zhang (2010)\cite{gon16}.

\begin{flushleft}\textbf{\large{3. Technique}}\end{flushleft}

\textbf{\large{3.1 Basics of LOESS}}
\vspace{0.3cm}

LOESS is a model independent non-parametric regression technique. In this technique we do not assume any  prior nor do we assume anything about the cosmological model. It is local as we use the local neighbourhood of each focal point in turn  to infer the global trend of data.

Our aim is to reconstruct  the function which can explain the behaviour of the dataset using LOESS. The  brief  methodology  to generate the reconstructed values is as follows: (For details see  ref.\cite{montial}).

1) We choose a subset having $n$ points out of the $N$ data points in the neighbourhood of our focal point $z_{i,0}$. The difference between the focal point $z_{i,0}$ and the farthest point of the neighbourhood ($max|z_{i,0}-z_j|$) where($j = 1,2,\ldots n$), is known as the span or bandwidth $h$ of the subset.

2) We next consider a weight function $w_{ij}$ such that it gives more weightage to the points closer to the  focal point and less weight to points far from the focal point. It is in the form of a kernel  function i.e. $w_{ij}= F[(z_j-z_{i,0})/h)]$.

3) We fit this subset of the data to a local polynomial $f(a,b,c)$ of first or second order using the weight function $w_{ij}$. We define a chi-square over this subset of data

\begin{equation}\label{loesschi2}
\chi^2_i=\sum_{j=1}^{n} w_{ij} (\eta^{obs}_{j}-f_j(a,b,c))^2
\end{equation}

4) On minimizing this $\chi^2_i$ , we obtain the best fit values as $a_0$, $b_0$ and $c_0$. Then the reconstructed value of $\eta^{obs}_i$ at focal point $z_{i,0}$ will be given by

\begin{equation}\label{reconstreta}
\widehat{\eta}_i= f_i(a_0,b_0,c_0)
\end{equation}

Repeating  this procedure for each point  in a given dataset, we obtain the corresponding reconstructed value at that point.  Thus, we see that to use  LOESS, we need  to fix the  bandwidth or smoothing parameter, weight function and the degree of the local polynomial.

\vspace{0.3cm}

\textbf{a) Bandwidth}

\vspace{0.2cm}

For smoothing, we have to select the neighbouring points  around the focus point. This is decided by the smoothing parameter, $\alpha$. This relates the optimal number of neighbourhood points ($n$) to the total number of data points ($N$) i.e., $n= \alpha N$ where $\alpha$ ranges  from $0$ to $1$.

The ideal way to select the optimum value of smoothing parameter is the  ``leave one out cross validation technique''\cite{fox}.

 The reconstructed value of $\eta^{obs}_i$ is obtained by removing the $i^{th}$ observation for a given value of $\alpha$. The reconstructed value of $\eta^{obs}$ is  denoted by $\widehat{\eta}_{-i}$ and this process is repeated for the whole dataset.
The cross validation function ($CV$) is given by
\begin{equation}
CV(\alpha)= \frac{1}{N}\sum_{i=1}^{N}(\eta^{obs}_i-{\widehat \eta _{-i}})^2
\end{equation}

  $CV(\alpha)$ can be thought of as a kind of mean squared error for different values of the smoothing parameter $\alpha$. We calculate  $CV(\alpha)$ for different values of $\alpha$. The value of $\alpha$ for which $CV(\alpha)$ is minimum is chosen to be the  optimal choice of $\alpha$ and used as the smoothing parameter for the further calculations.

\vspace{0.3cm}

\textbf{b) Weight function}

\vspace{0.2cm}

 The choice of weight function is governed by the guess that points close to each other may be more correlated than points which are far away. Using this conjecture, we need to  give more weight to observations that are closer to the focal point $z_{i,0}$. Following this logic  we choose a tricube weight function,

\begin{equation}\label{weightfn}
w_{ij} =
\left\{ \begin{array}{rcl}
(1-u_{ij}^3)^3 & \mbox{for}
& |u_{ij}|<1 \\ 0 & \mbox{for} & \text{  otherwise} \\
\end{array}\right.
\end{equation}

where $u_{ij}$ is defined as

\begin{equation}\label{uij}
u_{ij}= \frac{z_{i,0}-z_j}{h}
\end{equation}

Here $h= \text{max}|z_{i,0}-z_j|$ is the maximum distance between the point of interest and the $j^{\text{th}}$ element of its window. For other possible weight functions and their comparison see Ref. \cite{soh}.

\vspace{0.3cm}

\textbf{c) Degree of polynomial}

\vspace{0.2cm}

 The whole idea of LOESS is to fit the small subset of data using polynomials of a  low degree. A higher degree  polynomial in practice may increase computational cost without giving any significant  improvement in result.  Therefore the  polynomial fit for each subset of the data is usually of first or second degree i.e. linear or quadratic. In this work we fit neighbourhood subset of each focal point with a linear polynomial fit.

\begin{equation}
f_i(a,b)= a+bz_i
\end{equation}

\begin{figure*}
\includegraphics[height=4cm,width=5.85cm,scale=4]{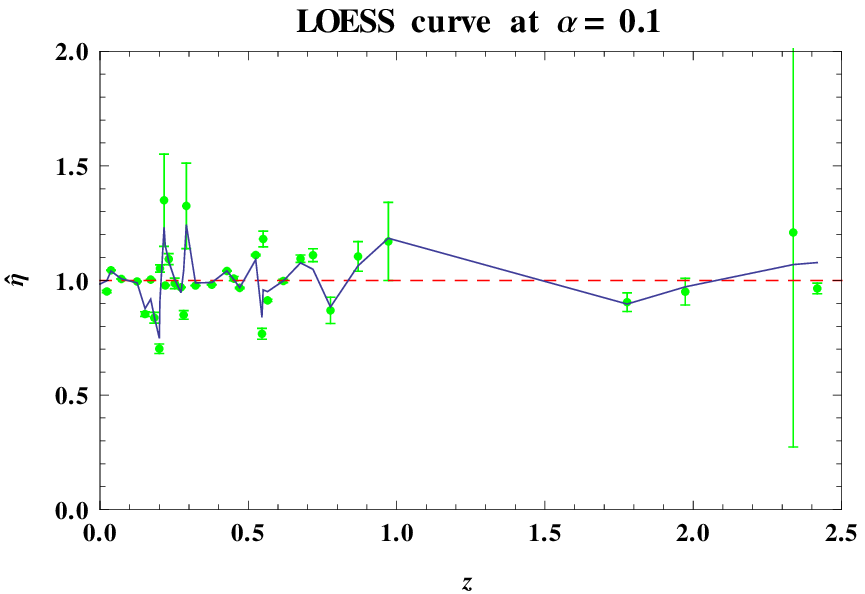}
\includegraphics[height=4cm,width=5.9cm,scale=4]{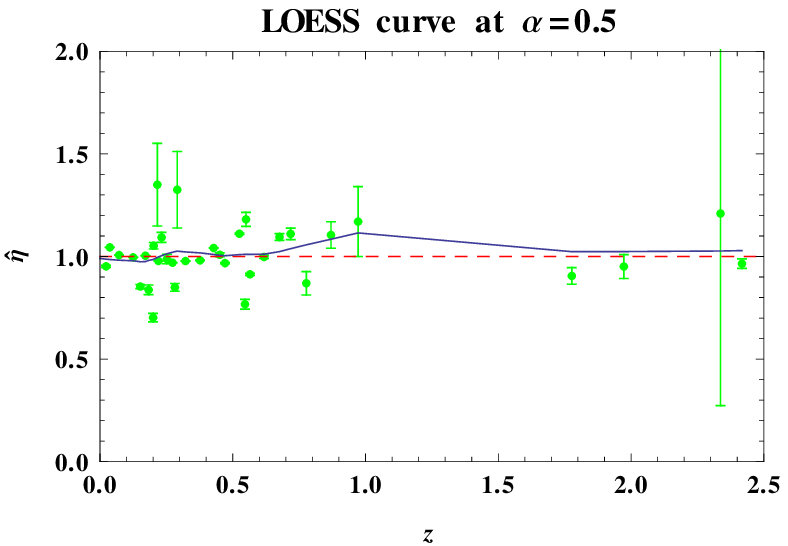}
\includegraphics[height=4cm,width=5.95cm,scale=4]{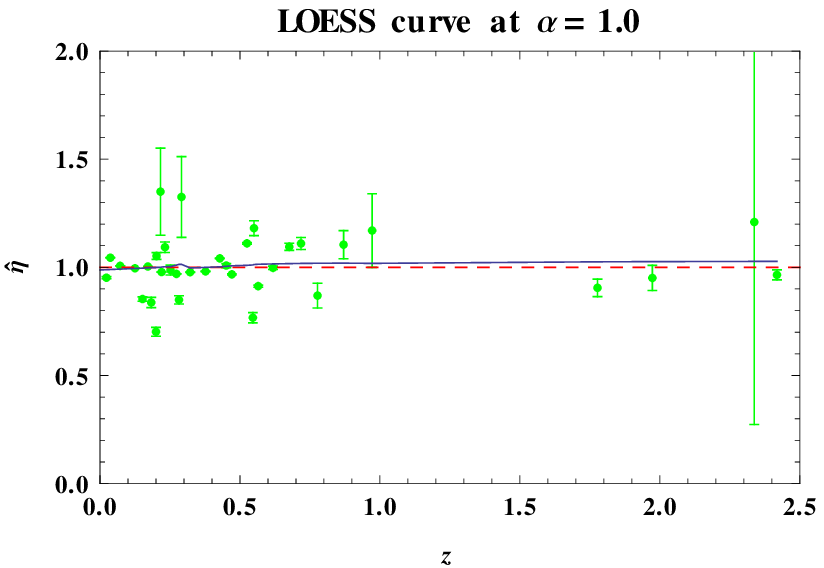}
\caption{LOESS plots  of $\widehat{\eta}$  with different bandwidhts ($\alpha=0.1,0.5,1.0$) for Dataset B. The red dashed line corresponds to $\widehat{\eta}=1$. }
\label{fig2}
\end{figure*}

\textbf{d) Confidence region around the reconstructed curve}

To construct the  $1 \sigma$ and $2 \sigma$ confidence regions around the nonparametric regression curve of $\widehat\eta$ , we assume that the errors are distributed normally \cite {fox,wasserman}. This can be obtained  by using the limiting values: $\widehat \eta_i\pm \sqrt{V(\widehat \eta_i)}$ and $\widehat \eta_i\pm 2\sqrt{V(\widehat \eta_i)}$ respectively, where

\begin{equation} \label{aa}
 V(\widehat{\eta}_i)= {1\over df_{res}}\sum_{i}^{N} d_i^2\sum_{j}^{N}w_{ij}^2
\end{equation}
where $d_i = \eta_i - \widehat\eta_i$ and  $df_{res}= N- df_{mod}$. $df_{mod}$ is the number of effective degrees of freedom or the effective number of parameters used in this regression. We calculate it by using   normalized smoothing matrix \textbf{S}. Then  $df_{mod}= Tr($\textbf{SS}$^T$). The smoothing matrix \textbf{S}, which is a square $ N \times N$ matrix of $w_{ij}$ elements,  is  directly calculated from the weight function. For more details see Refs \cite{montial,fox}.

\begin{figure*}

\includegraphics[height=4cm,width=5.5cm,scale=4]{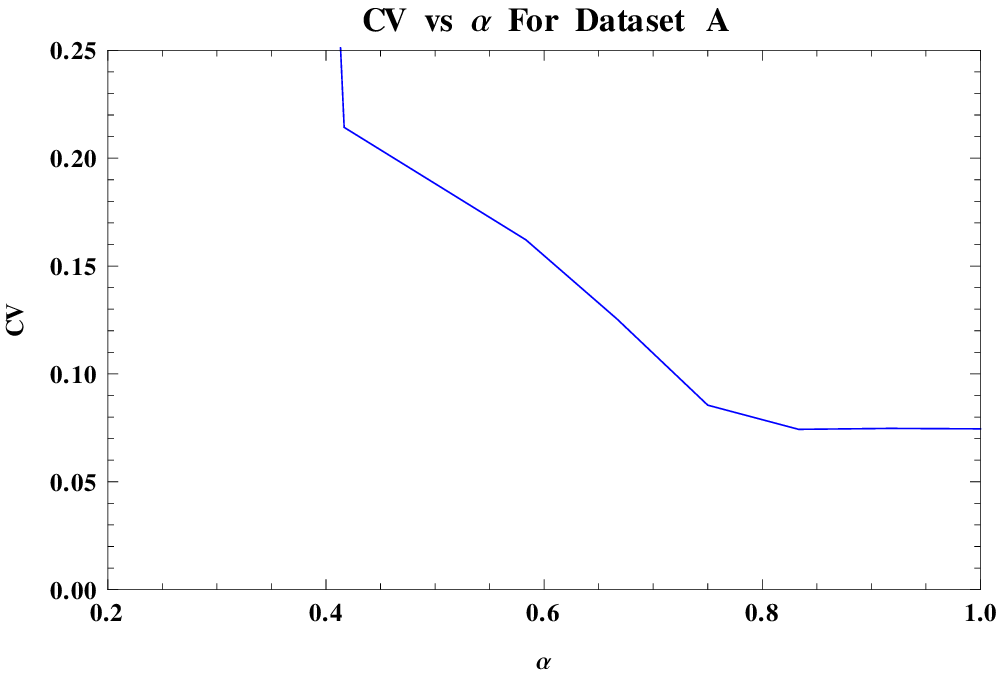}
\includegraphics[height=4cm,width=5.5cm,scale=4]{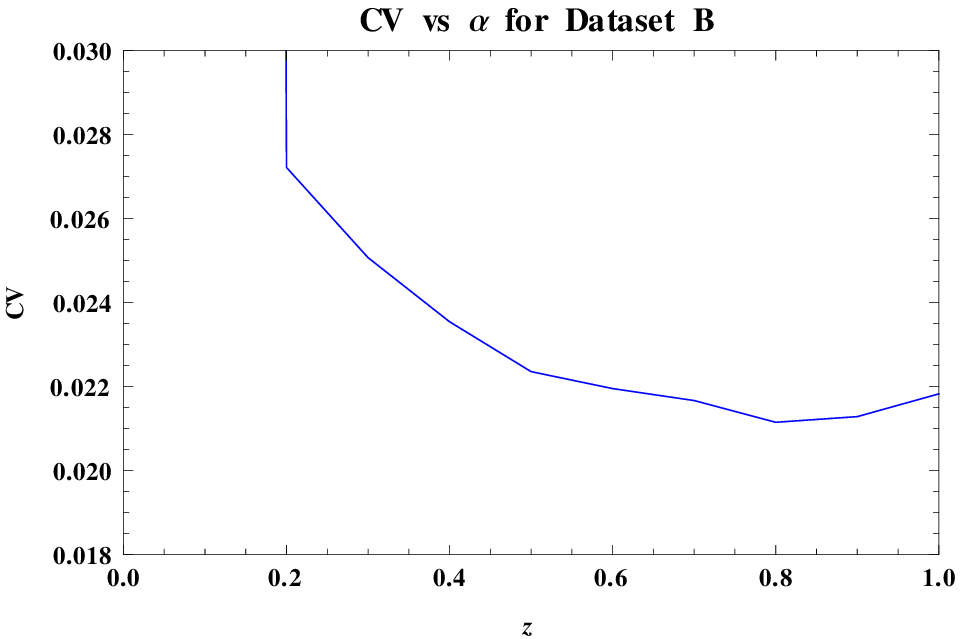}
\includegraphics[height=4cm,width=5.5cm,scale=4]{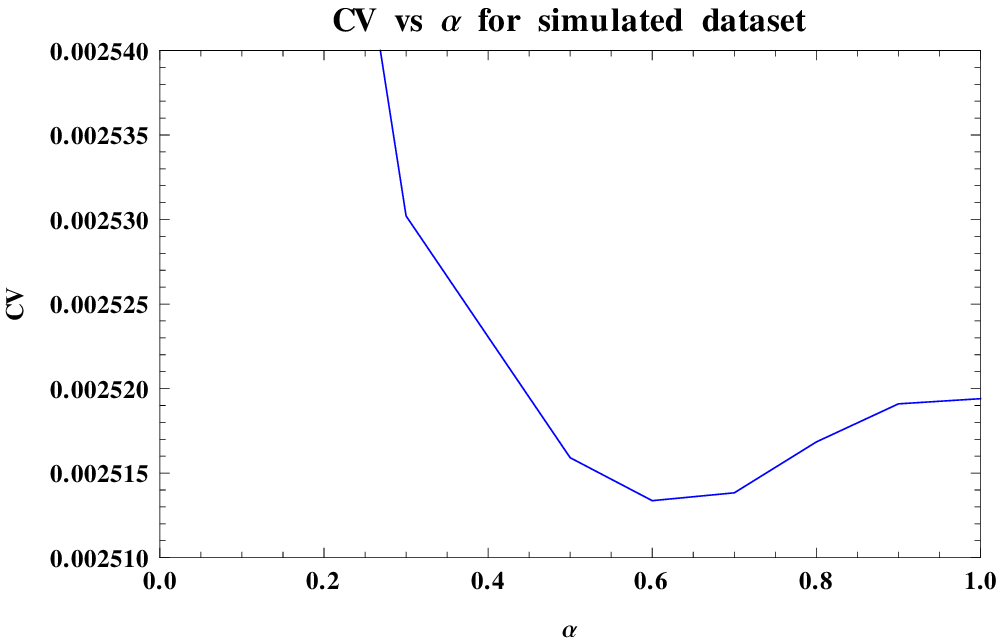}

\caption{Plot of $CV$ vs $\alpha$ for all the datasets. For Dataset A, we choose $\alpha$ to be 0.9 , while for Dataset B and Simulated dataset  minimum value of CV  is at $\alpha= 0.8$ and $\alpha= 0.6$ respectively.  }
\label{fig3}
\end{figure*}

\vspace{0.4cm}

\textbf{\large{3.2 Basics of SIMEX}}

\vspace{0.3cm}

In cosmology most of the observed quantities come with some degree of noise or measurement errors. However, in the LOESS method we don't  use the observed measurement  errors $\sigma_i$ while reconstructing the response parameter.
The effect of observational errors  can be accommodated in the  LOESS method by using a statistical technique called SIMEX. This was first introduced by Cook \& Stefanski (1994)  and then widely used in many fields for the purpose of reducing bias and measurement error from the data\cite{cook1 ,cook2}.

 SIMEX is based on a  two step resampling approach. Initially in the simulation step,  some  additional error is introduced by hand  in the data with some  controlling parameter $\xi$. Next by using regression analysis on this new  dataset, we try to trace the effect of  the measured error in our original dataset\cite{biewen , montial}.

1. For  the simulation step, we  introduce a fixed amount of measurement error in each observation data point  and define  a new variable as

\begin{equation}\label{simexeta}
\eta^{rec}_i(\xi_k)= \eta^{obs}_i + \sqrt{\xi_k} \,\,\,\sigma_i
\end{equation}
where $\sigma_i$ is the measurement error associated with the observed data $\eta^{obs}_i$. Here the  parameter $\xi_k$ acts as the controlling parameter for the variance of the measurement error. It is a vector of length $K$ such that $\xi_k>0$. We thus   form a matrix of order $K \times N$. The elements of this matrix will be the values of $\eta^{rec}_i(\xi_k)$ where $i= 1,2 \ldots N $ and $k= 1,2 \ldots K $. We chose  $\xi_k= 0.5,0.6,0.7 \ldots 2.0$ \cite{carroll,montial}.

\vspace{0.3cm}

2. We then  reconstruct  every data point given in each row of the matrix by applying the LOESS technique (as discussed in Section 3.1).
 Then by selecting each column of the reconstructed values of $\eta^{rec}$,  we can apply the simple regression technique to find the best fit value. We find that a quadratic polynomial is a better choice for this work.

Finally,  we obtain a row having $N$ elements. Each $\widehat{\eta}^{rec}_i(\xi_k)$ can be written as a function of $\xi_k$ i.e.
\begin{equation}\label{simexeta}
\widehat{\eta}^{rec}_i(\xi_k)= k_1+k_2 \xi_k+k_3 \xi_k^2
\end{equation}

If a normal distribution of the errors is assumed, then the error variance associated with the simulated data points $\widehat{\eta}^{rec}_i$ will be $(1+\xi_k)\sigma_i^2$. Substituting  $\xi_k=-1$, we are left with error-free smoothed data points.

\vspace{0.3cm}

\textbf{\large{4. Results}}

\vskip  0.2 cm
A critical point of the LOESS technique is the fixing of  the optimum value of the smoothing parameter ($\alpha$), i.e. the appropriate number of neighbourhood data points around the focus point. Fig.\ref{fig2} displays the LOESS curves for  different values of $\alpha$  for Dataset B. It is clear that for small values of the smoothing parameter (for instance $\alpha = 0.1$), the LOESS curve isn't smooth as it is very sensitive to random errors and fluctuations in the data. But as the span width increases, the smoothness of the curve increases. However, a high value of  $\alpha$ leads to over-smoothing \cite{montial}. The best way to find out the proper choice for $\alpha$  is the leave one out cross validation technique. We have to pick that value of $\alpha$  for which CV($\alpha$) is minimum. In Fig \ref{fig3},  we plot CV versus $\alpha$  for all three datasets. It is seen that for Dataset A the CV value decreases quickly at first and then shows no significant change after reaching the tail of the curve. Thus, any choice of $\alpha$  between $0.8$ and $1$, will be reasonable as the behaviour of $\widehat{\eta}$ remains the  same for the values of $\alpha \geq 0.8$ for the Dataset A.  However, to avoid the over smoothing of the curve  we will not choose $\alpha = 1$ [47]. Here  we choose $\alpha = 0.9$ to be the optimal choice of smoothing parameter for Dataset A. On the other hand, we can see that for Dataset B and the simulated dataset, we obtain a minimum value of CV at $\alpha = 0.8$ and $\alpha = 0.6$ respectively. 
 \begin{figure*}

\includegraphics[height=4cm,width=6cm,scale=4]{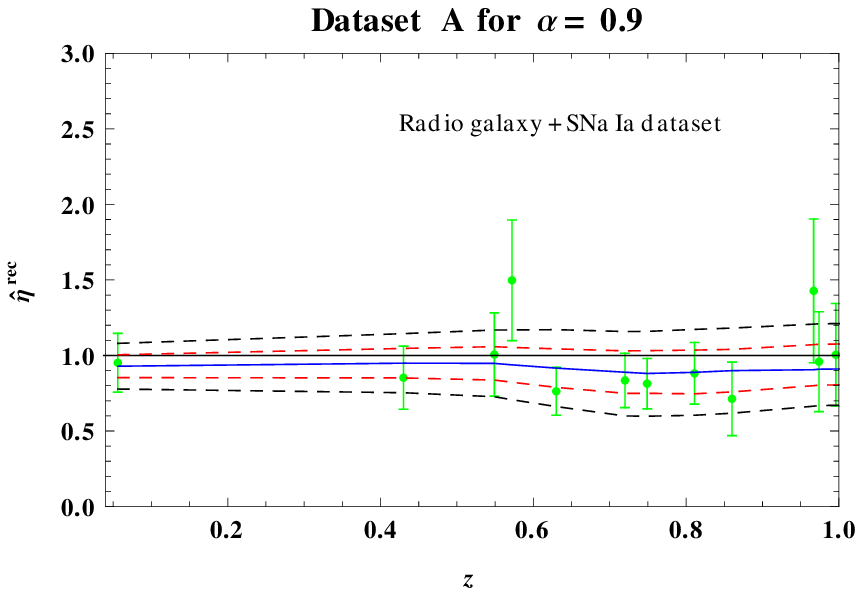}
\includegraphics[height=4cm,width=6cm,scale=4]{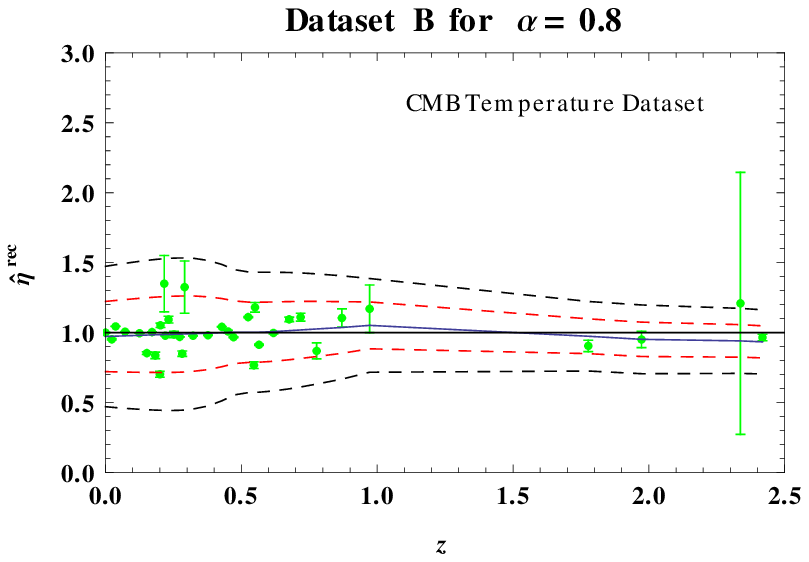}
\includegraphics[height=4cm,width=6cm,scale=4]{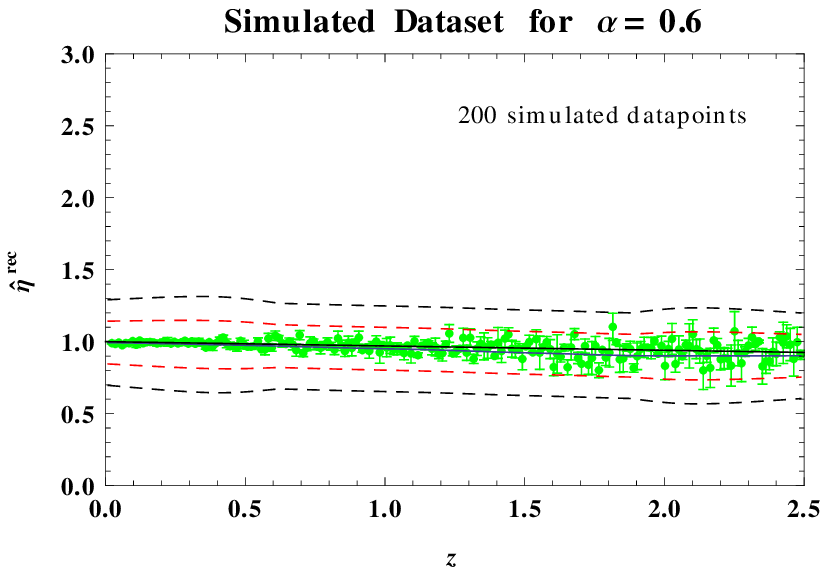}

\caption{\small LOESS + SIMEX  plots  of  $\widehat{\eta}^{rec}$ vs $z$  for  Dataset A (left), Dataset B(middle) and the simulated dataset (right). The dark blue line shows the reconstructed curve of $\widehat{\eta}^{rec}$ vs $z$. The red and black dash lines represent  $1 \sigma$ and $2 \sigma$ confidence levels for  best fits. Green points are the observed values. In Dataset A and Dataset B, solid black line represents $\widehat{\eta}^{rec}=1$  while in the simulated dataset, it shows the behaviour of fiducial model. }
\label{fig 4}
\end{figure*}

Fig.\ref{fig4} highlights the reconstruction of $\eta(z)$ along with the confidence regions by using LOESS + SIMEX  methods  for the  Dataset A , B and the mock (simulated) dataset respectively. The main features of our results are as follows:

1) For the simulated dataset, reconstructed $\eta(z)$ (using LOESS and SIMEX)  matches very well with the fiducial model. In addition, the $1 \sigma$ and $2 \sigma$ region accommodate all the data points and the fiducial line very well. This gives us confidence in the correctness and  efficiency of the methodology.

2) Dataset A is formed by using the  $D_A$ of radio galaxies. This data shows a very tight confidence band along the best fit line. It  supports DDR within the $1\sigma$ level for the redshift range $z< 1$.

 3) In Dataset B, we use the temperature of CMB radiation at various redshifts. The wide redshift range of this dataset makes it an important tool to study the behaviour of DDR even at very high redshift (upto  $z =2.418$). This is significant since this was not possible with  earlier datasets such  as galaxy clusters or BAO. On treating this dataset with  a non-parametric regression technique, we find that the reconstructed curve nearly overlaps the  $\eta = 1$ line in the entire redshift range within $ 1 \sigma$ level. In other words, there seems to be no  evidence of deviation from $ \eta = 1$.

 4) From these datasets, one can also infer that DDR is non-dynamical in nature as it stays close to $1$ in the entire redshift range used in this analysis.

\vskip 0.8 cm
\textbf{\large{5. Discussion}}  
\vskip 0.3 cm
The main objective of this work is to check the validity of the distance duality relation and its twin, the  temperature redshift relation by using a non-parametric method. We use the LOESS and SIMEX techniques to study the variation of $\eta(z)$ with  redshift. The main characteristic of this technique is that it does not require any prior or input from cosmological models. In order  to check the efficiency of this methodology we simulate a data set containing  $200$  equidistant data points of $\eta (z)$ in the redshift range $ 0< z \leq 2.418$ based on a fiducial model. In order to model the errors associated with this synthetic (simulated)  dataset,  we use the uncertainties associated  with the CMB temperature data at various redshifts.  By applying this methodology on the simulated dataset,  we find an excellent agreement between the reconstructed  and fiducial $\eta$ curves ( See Fig \ref{fig4}). This provides  further supportive evidence in the favour of this non-parametric regression technique.

Non-parametric methods have been used extensively in the literature to constrain DDR. Nesseris \& Bellido (2012) used  a non-parametric technique, namely  Genetic Algorithm (GA) with SNe Ia and BAO data and found that the  reconstructed curve is  compatible with $\eta=1$\cite{nesseries}.  Gaussian Process (GP) is another non-parametric technique used very frequently in cosmology for studying the variation of observational parameters\cite{shaif, hols , seikel}.
Zheng (2014) use the GP method with Union 2.1 SNe  Ia and galaxy cluster data with elliptical and spherical $\beta$ profiles. Their  analysis supports the elliptical morphology of galaxy clusters only \cite{zheng}. Nair et al.(2015) used BAO, $H(z)$ \& SNe Ia data to constrain DDR by the GP  method and found no evidence of deviation from $\eta(z) = 1$ \cite{remya1}. Recently Costa et al.(2015)   used the GP technique and applied it to  the two datasets based on galaxy cluster observations of angular diameter distance and gas mass fraction. They find that DDR is valid within the  $1\sigma$ region for the spherical $\beta$ model \cite{costa}.
Shafieloo et al.(2013) introduced a non-parametric smoothing method and crossing statistics to check whether any inconsistencies exist between the SNIa data and cluster data by assuming DDR to be valid \cite{sahni}.

All these techniques are quite effective but they need some a priori information (GP, NPS) to initiate. In addition, sometimes the lack of proper tools to estimate and propagate errors (GA) may lead to an underestimation of the  error. In this work, we don't assume any kind of prior information or any cosmological model and  by combining  LOESS with SIMEX , we believe that the   errors are also taken into account and are  not underestimated. Further, we think that this  methodology provides  an independent way to study the  validity of DDR.

Earlier work on DDR either used galaxy clusters or BAO for angular diameter distances. Recently Ma \& Corasaniti (2016) tested DDR with JLA SNe Ia along with $D_A$ estimates from BAO measurements of the WiggleZ Survey and BOSS DR12\cite{cong}. Interestingly, some authors use Gravitational lensing (distance ratio) data for $D_A$ and find no violation in DDR \cite{liao,ho}.

Finally, it is important to stress that no significant evidence of DDR  violation is observed in the entire redshift range $0<z\leq 2.418$ used in this analysis.
Further the luminosity distance $D_L$  used for the calculation of $\eta_{obs}$ in Dataset A, is obtained under the assumption of cosmic concordance ($\Lambda CDM$) model. Our analysis with Dataset A also highlights that DDR is compatible at $1 \sigma$ level in the $\Lambda CDM$  framework.
 Since this technique is data dependent, it is obvious that we will get  much better results as the quality and quantity  of data improves in the near future.

\vspace{0.3cm}
\begin{flushleft}
\textbf{\large{Acknowledgements: }}  Authors are thankful to anonymous referee for useful comments which substantially improve the paper.
A.R. acknowledges support under an CSIR-JRF scheme (Govt. of India, FNo. 09/045(1345/2014-EMR-I)). D.J. is thankful to CTP (Jamia Millia Islamia, New Delhi) for providing the facilities and research support. A.M. thanks Research Council, University of Delhi, Delhi for providing support under R \& D scheme 2015-16.
\end{flushleft}


\small
\begin{thebibliography}{1}

\bibitem {ethr} I.M.H. Etherington, \textit{On the Definition of Distance in General Relativity}, Phil. Mag. 15 (1933) 761.

\bibitem {ellis}  G.F.R. Ellis, \textit{Editorial note on "On the definition of distance in general relativity by I. M. H. Etherington "}, Gen. Rel. Grav. 39 (2007) 1047.

\bibitem{lima1} J.A.S. Lima, J.V. Cunha \& J.S. Alcaniz, \textit{
	Constraining the dark energy with galaxy cluster x-ray data}, Phys. Rev. D 68 (2003) 023510.

\bibitem{lima2} J.V. Cunha, L. Marassi \& J.A.S.Lima, \textit{Constraining H0 from the Sunyaev-Zel'dovich effect, galaxy cluster X-ray data and baryon oscillations}, Mon. Not. R. Astron. Soc. 379 (2007)  L1.

\bibitem{schneider} P.Schneider, J.Ehlers \& E.E.Falco, \textit{ Gravitational Lenses }, ISBN 978-3-662-03758-4 (Springer-1999).

\bibitem{komatsu} E. Komatsu et al., \textit{Seven-Year Wilkinson Microwave Anisotropy Probe (WMAP) Observations: Cosmological Interpretation}, Astrophys J. Suppl. Ser. 192 (2011) 18.

\bibitem{basset} B.A. Basset \& M. Kunz, \textit{Cosmic distance-duality as a probe of exotic physics and acceleration}, Phys Rev D 69 (2004) 101305.

\bibitem{rfl} R.F.L. Holanda, R.S. Goncalves \& J.S. Alcaniz,\textit{ A test for cosmic distance duality}, J. Cosmol. Astropart. Phys. 06 (2012) 022.

\bibitem{goncalves} R.S. Goncalves, J.S. Alcaniz, J.C. Carvalho \& R.F.L. Holanda, \textit{Forecasting constraints on the cosmic duality relation with galaxy clusters }, Phys. Rev. D  91 (2015) 027302

\bibitem{goncalves22}  R.S. Goncalves, A. Bernui, R.F.L. Holanda \&  J.S. Alcaniz, \textit{Constraints on the duality relation from ACT cluster data }, Astron. \& Astrophys. 573 (2015) A88.

\bibitem{gfr} G.F.R. Ellis, R. Poltis, J.P. Uzan \& A. Weltman, \textit{The blackness of the cosmic microwave background spectrum as a probe of the distance-duality relation }, Phys. Rev. D. 87 (2013) 103530.

\bibitem{ruth} R. Lazkoz, S. Nesseris \& L. Perivolaropoulos, \textit{Comparison of Standard Ruler and Standard Candle constraints on Dark Energy Models }, J. Cosmol. Astropart. Phys. 07 (2008) 12.

\bibitem{sysky} S. Rasanen, J. Valiviita \& V. Kosonen, \textit{Testing distance duality with CMB anisotropies }, J. Cosmol. Astropart. Phys. 04 (2016) 50.

\bibitem{rfl2} R.F.L. Holanda \& V.C. Busti, \textit{Probing cosmic opacity at high redshifts with gamma-ray bursts }, Phys. Rev. D 89 (2014) 103517.





\bibitem{khedekar} S. Khedekar \& S. Chakraborti, \textit{A new Tolman test of a cosmic distance duality relation at 21 cm }, Phys. Rev. Lett. 106  (2011) 221301.

\bibitem{cong} C. Ma \& P.S. Corasaniti, \textit{Statistical Test of Distance-Duality Relation with Type Ia Supernovae and Baryon Acoustic Oscillations }, (2016) [arXiv:astro-ph/1604.04631].

\bibitem{cardone} V.F. Cardone, S. Spiro, I. Hook \& R. Scaramella, \textit{Testing the distance duality relation with present and future data}, Phys. Rev. D 85 (2012) 123510.

 \bibitem{more} S. More, J. Bovy \& D.W. Hogg , \textit{Cosmic transparency: A test with the baryon acoustic feature and type Ia supernovae}, Astrophys. J.  696 (2009) 1727.

\bibitem{avgo} A. Avgoustidis, L. Verde \& R. Jimenez, \textit{Consistency among distance measurements: transparency, BAO scale and accelerated expansion}, J. Cosmol. Astropart. Phys. 06 (2009) 012.

\bibitem{nicole} H. Lampeitl, R. C.Nichol et al.,\textit{First-year Sloan Digital Sky Survey-II (SDSS-II) supernova results: consistency and constraints with other intermediate-redshift datasets}, Mon. Not. R. Astron. Soc. 401 (2009) 2331.

\bibitem{djain}  R. Nair, S. Jhingan \& D. Jain, \textit{Cosmic distance duality and cosmic transparency}, J. Cosmol. Astropart. Phys. 12 (2012) 028.

\bibitem{csaki} 	C. Csaki, N. Kaloper \& J. Terning,  \textit{Dimming Supernovae without Cosmic Acceleration }, Phys. Rev. Lett. 88 (2002) 16.

	
\bibitem{jas}	J.A.S. Lima, J.V. Cunha \& V.T. Zanchin, \textit{Deformed Distance Duality Relations and Supernovae Dimming},  Astrophys. J. 742 (2011) 2.

\bibitem{suto} Y. Inagaki, T. Suginohara, \& Y. Suto, \textit{Reliability of the Hubble-Constant Measurement Based on the Sunyaev-Zel'dovich Effect}, Pub. of  Astronom. Soc. of Japan  47 (1995) 411.

\bibitem{pier} P. S. Corasaniti,\textit{The Impact of Cosmic Dust on Supernova Cosmology }, Mon. Not. R. Astron. Soc. 372 (2006) 191.

\bibitem{mirizzi} A. Mirizzi, G. G. Raffelt \& P. D. Serpico, \textit{Photon-axion conversion in intergalactic magnetic fields and cosmological consequences }, Lect. Notes Phys. 741 (2008) 115.

\bibitem{mincos} F. Piazza \& T, Schucker, \textit{Minimal cosmography },  Gen. Rel. Grav. 48 (2016) 41.

\bibitem{marek}  A. Piorkowska \& M. Biesiada, \textit{Distance Duality in Different Cosmological Models}, ACTA Physica Polonica B 42 (2011) 11.

\bibitem{remya} R. Nair, S. Jhingan \& D. Jain, \textit{Cosmokinetics: A joint analysis of Standard Candles, Rulers and Cosmic Clocks }, J. Cosmol. Astropart. Phys. 05 (2011) 023.

\bibitem{holanda} R.F.L. Holanda, J.A.S. Lima \& M.B. Ribeiro, \textit{Testing the Distance-Duality Relation with Galaxy Clusters and Type Ia Supernovae },  Astrophys J. 722 (2010) L233.

\bibitem{holandaI} R.F.L. Holanda, J.A.S. Lima \& M.B. Ribeiro, \textit{Probing the Cosmic Distance Duality Relation with the Sunyaev-Zeldovich Effect, X-rays Observations and Supernovae Ia }, Astron. \& Astrophys. 538 (2012) A131.

\bibitem{wu} Z. Li, P. Wu \& H. Yu, \textit{Cosmological-model-independent Tests for the Distance-Duality Relation from Galaxy Clusters and Type Ia Supernova}, Astrophys J. 729 (2011) L14.

\bibitem{meng} X.L. Meng, T.J. Zhang, H. Zhan  \&  X. Wang, \textit{ 	Morphology of Galaxy Clusters: A Cosmological Model-Independent Test of the Cosmic Distance-Duality Relation}, Astrophys J. 745 (2012) 98.

\bibitem{liang} S. Cao and N.Liang, \textit{Interaction between dark energy and dark matter: observational constraints from OHD, BAO, CMB and SNe Ia }, Res. Astron. \& Astrophys.  11 (2011) 1199.





\bibitem{yang} X. Yang, H.R. Yu, Z.S. Zhang \& T.J. Zhang,\textit{An improved method to test the Distance--Duality relation }, Astrophys J. 777 (2013) L24.

\bibitem{costa} S.S. Costa, V.C. Busti \& R.F.L. Holanda, \textit{Two new tests to the distance duality relation with galaxy clusters},  J. Cosmol. Astropart. Phys. 01 (2015) 061.

 \bibitem{sahni} A. Shafieloo, S. Majumdar, V. Sahni \& A.A. Starobinsky,
 \textit{Searching for systematics in SNIa and galaxy cluster data using the cosmic duality relation},  J. Cosmol. Astropart. Phys. 04 (2013) 042.

\bibitem{remya1} R. Nair, S. Jhingan, \& D. Jain,\textit{Testing the consistency between cosmological measurements of distance and age}, Phys. Lett. B 745 (2015) 64.

 \bibitem{zheng} Yi Zheng, \textit{Reconstruct the Distance Duality Relation by Gaussian Process}, (2014) [arXiv:astro-ph/1408.3897]


\bibitem{nesseries} S. Nesseris \& J.G. Bellido, \textit{A new perspective on Dark Energy modeling via Genetic Algorithms}, J. Cosmol. Astropart. Phys. 11 (2012) 033.

\bibitem{cleveland1} W. Cleveland, \textit{Robust Locally Weighted Regression and Smoothing Scatterplots}, J. Amer. Statist. Assoc. 74 (1979)  829.

\bibitem{cleveland2} W. Cleveland \& S. J. Devlin, \textit{Locally Weighted Regression: An Approach to Regression Analysis by Local Fitting}, J. Amer. Statist. Assoc. 83 (1988) 596.

\bibitem{cook1} J. R. Cook \& L. A. Stefanski, \textit{Simulation-Extrapolation Estimation in Parametric Measurement Error Models}, J. Amer. Statist. Assoc. 89 (1994) 1314.

\bibitem{cook2} L. Stefanski \& J. Cook, \textit{Simulation-extrapolation: The measurement error jackknife}, J. Amer. Statist. Assoc. 90 (1995)
1247.

\bibitem{biewen} E. Biewen , S. Nolte \& M. Rosemann , \textit{Multiplicative Measurement Error and the Simulation Extrapolation Method}, IAW Discussion Paper  39 (2008).

\bibitem{montial} A. Montiel, R. Lazkoz, I. Sendra, C. E. Rivera \& V. Salzano, \textit{Nonparametric reconstruction of the cosmic expansion with local regression smoothing and simulation extrapolation}, Phys. Rev. D. 89 (2014) 043007.
	
\bibitem{fabris} C.E. Rivera, J.C. Fabris \& C.Julio, \textit {Nonparametric reconstruction of the Om diagnostic to test LCDM}, (2015), [arXiv:astro-ph/1511.07066].

\bibitem{nisha} N. Rani, D. Jain, S. Mahajan, A. Mukherjee \& N. Pires,
   \textit{Transition redshift: new constraints from parametric and nonparametric methods}, J. Cosmol. Astropart. Phys. 12 (2015) 045.
	
\bibitem{betoule} M. Betoule, R. Kessler \& J. Guy et.al. \textit{Improved cosmological constraints from a joint analysis of the
SDSS-II and SNLS supernova samples.}, Astron. \& Astrophys 568 (2014) A22.




\bibitem{routhdaly}  R. A. Daly \& S.G. Djorgovski, \textit{A Model-Independent Determination of the Expansion and Acceleration Rates of the Universe as a Function of Redshift and Constraints on Dark Energy }, Astrophys. J. 597 (2003) 9


\bibitem{cao} S. Cao \& N. Liang, \textit{Testing the distance-duality relation with a combination of cosmological distance observations}, Research in Astronomy and Astrophysics, 11 (2011) 10.

\bibitem{baranov}I. Baranov, J.F. Jesus \& J.A.S. Lima, \textit{Testing CCDM Cosmology with the Radiation Temperature-Redshift Relation }, (2016)  [arXiv:astro-ph/1605.04857],



\bibitem{gon16} Cong Ma \& T. Zhang , \textit{Power of Observational Hubble Parameter Data: a Figure of Merit Exploration}, Astrophys.J. 730 (2011) 74.



\bibitem{fox} J. Fox, \textit{Nonparametric Simple Regression: Smoothing
Scatterplots}, ISBN 978-0-761-91585-0 (SAGE Publications, 2000).

\bibitem{soh}  Y. Soh, Y. Hae, A. Mehmood, R.H. Ashraf \& I. Kim, \textit{Performance Evaluation of Various Functions for Kernel Density Estimation}, Open J. App. Sciences 3 (2013) 58.

\bibitem{wasserman} L. Wasserman, \textit{All of Nonparametric Statistics}, ISBN 978-0-387-30623-0 (Springer, 2006).

\bibitem{carroll} T. V. Apanasovich, R. J. Carroll \& A. Maity, \textit{SIMEX and standard error estimation
in semiparameteric measurement error}, Electron.J. Statist. 3 (2009) 318.



\bibitem{shaif} A. Shafieloo et al.,\textit{Gaussian Process Cosmography }, Phys. Rev. D 85 (2012) 123530.

\bibitem{hols} T. Holsclaw, U. Alam, B. Sanso et al., \textit{Nonparametric Dark Energy Reconstruction from Supernova Data }, Phys. Rev. Lett. 105 (2010) 241302.

\bibitem{seikel} M. Seikel, C. Clarkson \& M. Smith, \textit{Reconstruction of dark energy and expansion dynamics using Gaussian processes }, J. Cosmol. Astropart. Phys. 06 (2012) 036.

\bibitem{liao} K. Liao, Z. Li, S. Cao, M. Biesiada, X. Zheng \& Z.H. Zhu,  \textit{The Distance Duality Relation from Strong Gravitational Lensing }, Astrophys. J. 822 (2016) 74.

\bibitem{ho} R.F.L. Holanda, V.C. Busti \& J. S. Alcaniz, \textit{Probing the cosmic distance duality with strong gravitational lensing and supernovae Ia data }, J. Cosmol. Astropart. Phys. 02 (2016) 054.

\end{thebibliography}
\end{document}